\documentclass[reprint,superscriptaddress,secnumarabic,amssymb, nobibnotes,noeprint, aps, prd,numerical]{revtex4-2}

\usepackage{graphicx}
\graphicspath{{images/}} 
\usepackage{siunitx}
\usepackage{xcolor}
\usepackage[colorlinks = true, linkcolor = black, urlcolor = black, citecolor = black, anchorcolor = black]{hyperref}

\begin{document}

\title{Current-phase relation of a WTe$_2$ Josephson junction}%

\author{Martin Endres}
\thanks{These authors contributed equally. }
\affiliation{Department of Physics, University of Basel, Klingelbergstrasse 82, 4056 Basel, Switzerland}

\author{Artem Kononov}
\thanks{These authors contributed equally. }
\affiliation{Department of Physics, University of Basel, Klingelbergstrasse 82, 4056 Basel, Switzerland}

\author{Hasitha Suriya Arachchige}
\affiliation{Department of Physics and Astronomy, University of Tennessee, Knoxville, Tennessee 37996, USA}

\author{Jiaqiang Yan}
\affiliation{Department of Physics and Astronomy, University of Tennessee, Knoxville, Tennessee 37996, USA}
\affiliation{Material Science and Technology Division, Oak Ridge Laboratory,Oak Ridge, Tennessee 37831, USA}

\author{David Mandrus}
\affiliation{Department of Materials Science and Engineering, University of Tennessee, Knoxville, Tennessee 37996, USA}
\affiliation{Department of Physics and Astronomy, University of Tennessee, Knoxville, Tennessee 37996, USA}
\affiliation{Material Science and Technology Division, Oak Ridge Laboratory,Oak Ridge, Tennessee 37831, USA}

\author{Kenji Watanabe}
\affiliation{Research Center for Functional Materials, National Institute for Materials Science, 1-1 Namiki, Tsukuba 305-0044, Japan}

\author{Takashi Taniguchi}
\affiliation{International Center for Materials Nanoarchitectonics, National Institute for Materials Science,  1-1 Namiki, Tsukuba 305-0044, Japan}

\author{Christian Sch{\"o}nenberger}
\email{christian.schoenenberger@unibas.ch}
\affiliation{Department of Physics, University of Basel, Klingelbergstrasse 82, 4056 Basel, Switzerland}
\affiliation{Swiss Nanoscience Institute, University of Basel, Klingelbergstrasse 82, 4056 Basel, Switzerland}

\date{\today}%

\begin{abstract}
When a topological insulator is incorporated into a Josephson junction, the system is predicted to reveal the fractional Josephson effect with a 4$\pi$-periodic current-phase relation. 
Here, we report the measurement of a $4\pi$-periodic switching current through an asymmetric SQUID, formed by the higher-order topological insulator WTe$_2$. Contrary to the established opinion, we show that a high asymmetry in critical current and negligible loop inductance are not sufficient by themselves to reliably measure the current-phase relation. Instead, we find that our measurement is heavily influenced by additional inductances originating from the self-formed PdTe$_{\text{x}}$ inside the junction. We therefore develop a method to numerically recover the current-phase relation of the system and find the \SI{1.5}{\micro \meter} long junction to be best described in the short ballistic limit. Our results highlight the complexity of subtle inductance effects that can give rise to misleading topological signatures in transport measurements.

\end{abstract}

\maketitle

\section*{Introduction}
Topological insulators (TIs) belong to a unique class of materials that are insulating in their bulk while hosting gapless boundary states that are protected by time-reversal symmetry \cite{Fu2007}. The class of three-dimensional TIs has recently been extended \cite{Schindler2018, Benalcazar2017}, realizing that a \textit{d}-dimensional TI of order \textit{n} can develop (\textit{d}-\textit{n})-dimensional hinge- or corner-states.  A promising candidate of this novel material class that is predicted to host topological hinge-states is the semi-metallic transition-metal-dichalcogenide WTe$_2$ \cite{Wang2019, Peng2017,Huang2020, Kononov2020}.  While bulk states dominate transport in the normal state, hinge states become the governing transport channel over long distances in the superconducting state as they can carry a higher critical current due to their reduced dimensionality \cite{Heikkila2002, Murani2017}.
Therefore, Josephson junctions (JJs) formed by a TI as the weak-link in between two superconducting electrodes provide an ideal platform to probe the topological nature in a transport experiment. Topological hybrid systems are of great interest as they are predicted to host unconventional superconductivity \cite{Fu2008}, the fundamental building block of a potential topologically protected quantum bit \cite{Hyart2013}.

The fingerprint of a JJ is the current-phase relation (CPR), the dependence of the supercurrent on the phase-difference $\varphi$ between the superconducting electrodes.  The measurement of the CPR directly reflects the underlying transport mechanism with which the Cooper pairs are shuttled across the weak link. For a topological weak-link, perfect Andreev reflection is expected \cite{Adroguer2010} since spin-momentum locking in the hinge-states prohibits normal electron reflection at the interface to the superconductor. 
In the long junction limit, the supercurrent is carried by 4$\pi$-periodic Andreev bound states of opposite parity, $\sigma^{+}$ and $\sigma^{-}$, that give rise to a characteristic sawtooth-shaped $I_{\text{c}}$. Parity conservation prohibits the recombination to the lower energy branch and results in a multivalued $I_{\text{c}}$ with a distinct diamond shape, as plotted in Fig.~\ref{fig:1}~a) \cite{Beenakker2013}. For comparison, the expected CPR of a trivial ballistic JJ in the long junction limit is plotted as black dashed line in the same figure. 
Accordingly, the literature often infers ballistic topological states from the observation of a sawtooth-shaped flux dependence of the critical current \cite{Kayyalha2020, Li2019, Murani2017, Bernard2023}. 

Recently, high quality JJs formed in WTe$_2$ on palladium (Pd) bottom contacts have been reported \cite{Kononov2021, Endres2022} and provided evidence of a non-sinusoidal CPR~\cite{Kononov2020, Choi2020}.  
Here, we combine such JJs based on Pd-induced superconductivity in WTe$_2$ with external superconducting leads, as illustrated in Fig.~\ref{fig:1}~b). 
A superconducting quantum interference device (SQUID) formed out of two such JJs is expected to reflect the CPR, provided that the critical current amplitudes of the two JJs are highly different and the loop inductance is negligible \cite{DellaRocca2007, Nichele2020}. Based on this method, we observe a switching current distribution with 4$\pi$-periodicity that resembles a topological JJ \cite{Beenakker2013}. Contrary to the topological interpretation, we provide an alternative explanation based on inductance effects \cite{Lefevre-Seguin1992}, which could be relevant for a number of previous experiments \cite{Kayyalha2020, Li2019, Murani2017, Bernard2023}. Importantly, the inductance contribution originates from the JJs themselves and exceeds the loop inductance. We further developed a numerical model based on the maximization of the supercurrent in the SQUID loop that allows us to exclude screening effects from the additional inductances and recover the real CPR. The calculations suggest that the critical current of our \SI{1.5}{\micro \meter} long junction is best reproduced in the short ballistic limit, despite its long physical length.

\section*{SQUID Measurements}

\begin{figure}[t]
    \centering
    \includegraphics[width=1\linewidth]{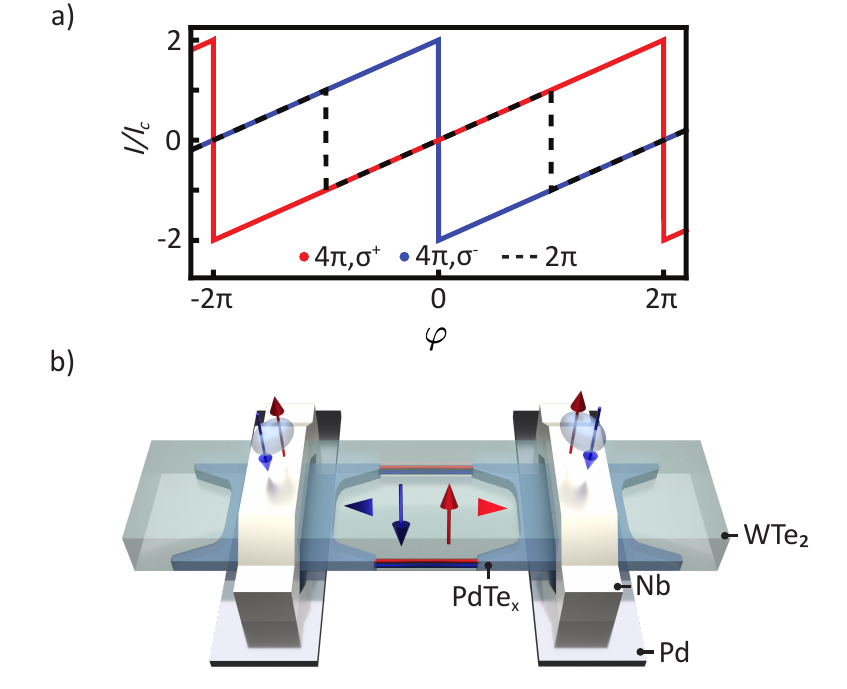}
    \caption{\textbf{CPR of a topological JJ.} a) Normalized CPR of a topological JJ in the long junction limit as a function of the junction phase $\varphi$. The 4$\pi$-periodic supercurrent through the SQUID is carried by Andreev bound states of opposite parity, $\sigma^{+}$ and $\sigma^{-}$, resulting in a multivalued switching current that resembles a diamond-like pattern.  A 2$\pi$-periodic CPR of a topologically trivial junction is shown as dashed line for comparison. b) llustration of a JJ fabricated from an elongated WTe$_2$ crystal on top of Pd bottom contacts. Superconducting Nb contacts were deposited from above after etching through the top hBN (not displayed).  The diffusion of Pd into the WTe$_2$ crystal leads to the formation of superconducting PdTe$_{\text{x}}$ inside the weak link, through which the topological hinge states can be coupled.}
\label{fig:1}
\end{figure} 

\begin{figure}[t]
    \centering
    \includegraphics[width=1\linewidth]{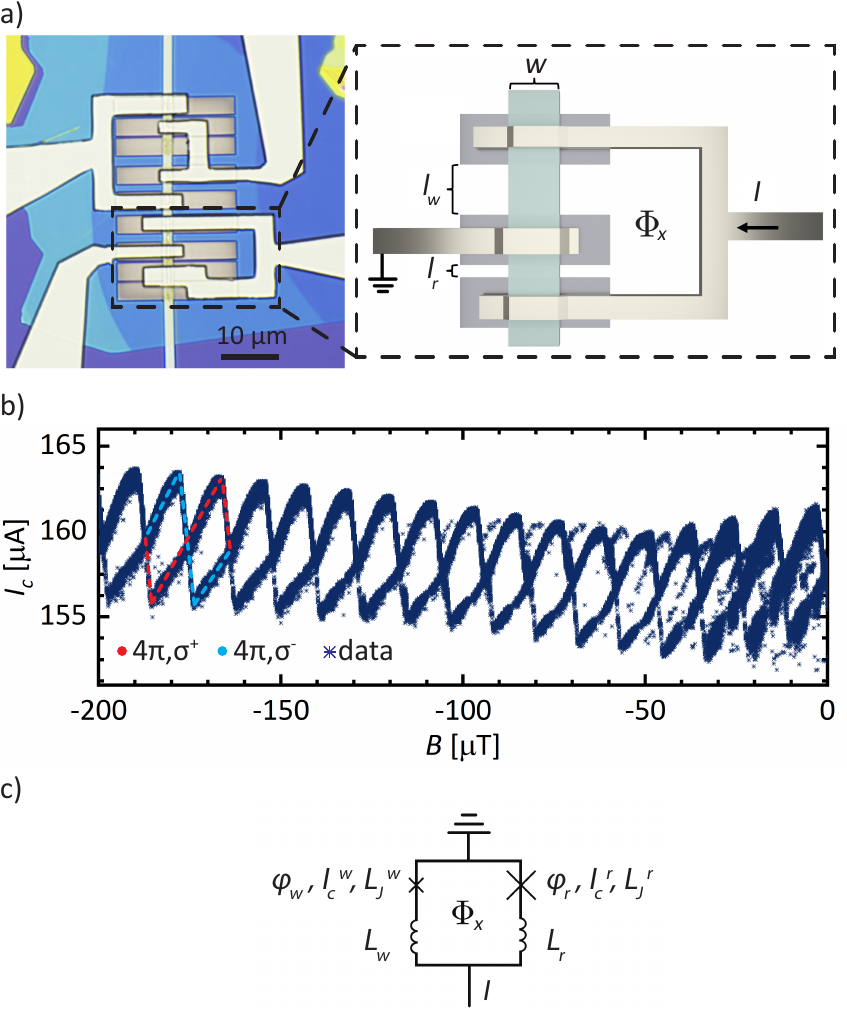}
    \caption{\textbf{Switching statistics of the weak junction in an asymmetric SQUID. } a) Optical image of SQUID device on the left,  with the measured SQUID highlighted by the dashed box. An illustration of the device parameters is provided to the right. b) High-resolution measurement of the SQUID switching current as a function of applied magnetic field through the loop. An oscillation period of the parity states $\sigma^{+}$ and $\sigma^{-}$, respectively, is highlighted in red and light blue, respectively. The period $\delta B=$ \SI{23.2}{\micro \tesla} of a single parity branch displays a $4\pi$ periodicity with respect to the designed loop area $\delta B=$ \SI{11.1}{\micro \tesla}. c) Schematic of the SQUID, specifying the device parameters, including the additional series inductances $L_{\text{w}}$ and $L_{\text{r}}$ in series to the weak and reference JJ, respectively.}%
    \label{fig:2}
\end{figure}

We begin with the fabrication of an asymmetric SQUID out of WTe$_2$.
In our experiment, both JJs are formed in the same needle-shaped WTe$_2$ flake of width $w=$ \SI{1.5}{\micro \meter}. The asymmetry in critical current $I_{\text{c}}$ of the two involved JJs is achieved by a different spacing between the Pd bottom contacts $l_{\text{w}}$ = \SI{1.5}{\micro \meter} and $l_{\text{r}}$ = \SI{0.5}{\micro \meter} for the weak and reference junction, respectively,  as sketched in Fig.~\ref{fig:2}~a). 
The superconducting loop is formed by etching through the top hBN into WTe$_2$ and sputtering niobium (Nb) on top. 
The Nb leads are between \SI{2.2}{\micro \meter} and \SI{3}{\micro \meter} wide and \SI{100}{\nano \meter} thick. 
A detailed description of the fabrication process can be found in the supplementary materials \cite{supp}.  Figure \ref{fig:2}~a) displays an optical image of the finished devices. In the following we will focus on the lower SQUID, enclosed by the dashed line. 

The critical current of the SQUID $I_{\text{c}}(\varphi_{\text{w}}, \varphi_{\text{r}})=I_{\text{c}}^{\text{w}}f_{\text{w}}(\varphi_{\text{w}})+I_{\text{c}}^{\text{r}}f_{\text{r}}(\varphi_{\text{r}})$ is the sum of the individual currents $I_{\text{w}}$ and $I_{\text{r}}$ through the two branches of the loop,  defined by the critical current $I_{\text{c}}^{\text{i}}$ and the normalized CPR $f_{\text{i}}$ of the $i^{\text{th}}$ Josephson junction. The total flux $\Phi_{\text{tot}}$ threading the loop connects the phase differences across the two JJs $\varphi_{\text{w}}-\varphi_{\text{r}}=2\pi\Phi_{\text{tot}}/\Phi_{0}=\phi_{\text{tot}}$, with $\phi_{\text{tot}}$ denoting the external phase. In the absence of inductances in the loop, $\phi_{\text{tot}}$ is simply defined by the external phase $\phi_{\text{x}}$.
The asymmetry $I_{\text{c}}^{\text{r}}\gg I_{\text{c}}^{\text{w}}$, pins $\varphi_{\text{r}} = \varphi_{\text{r}}^{\text{max}}$ at a fixed phase, for which $I_{\text{c}}^{\text{r}}$ is maximized \cite{DellaRocca2007, Indolese2020, Nichele2020}. The normalized CPR of the weak junction $f_{\text{w}}$ can then be deduced from the measurement of
\begin{equation}
I_{\text{c}}(\phi_{\text{x}})\sim I_{\text{c}}^{\text{w}}f_{\text{w}}(\varphi_{\text{r}}^{\text{max}}+\phi_{\text{x}})+I_{\text{c}}^{\text{r}}f_{\text{r}}(\varphi_{\text{r}}^{\text{max}}).
\label{Equ:1}
\end{equation}
In the experiment, $\Phi_{\text{x}} = BA_{\circ}$ is controlled by an applied perpendicular magnetic field $B$ threading the effective loop area $A_{\circ} =$ \SI{186}{\micro\meter\squared}. The Meissner screening of the enclosing superconducting loop was taken into account by including half of its width into the loop area.
The final device is probed in a quasi four-terminal configuration by sourcing an ac-current and monitoring the voltage drop over the SQUID.  We use the counter technique, as described in Ref.~\cite{supp}, to measure the switching statistics of the device.  

\begin{figure*}[t!]
   \centering
   \includegraphics[width=1\linewidth]{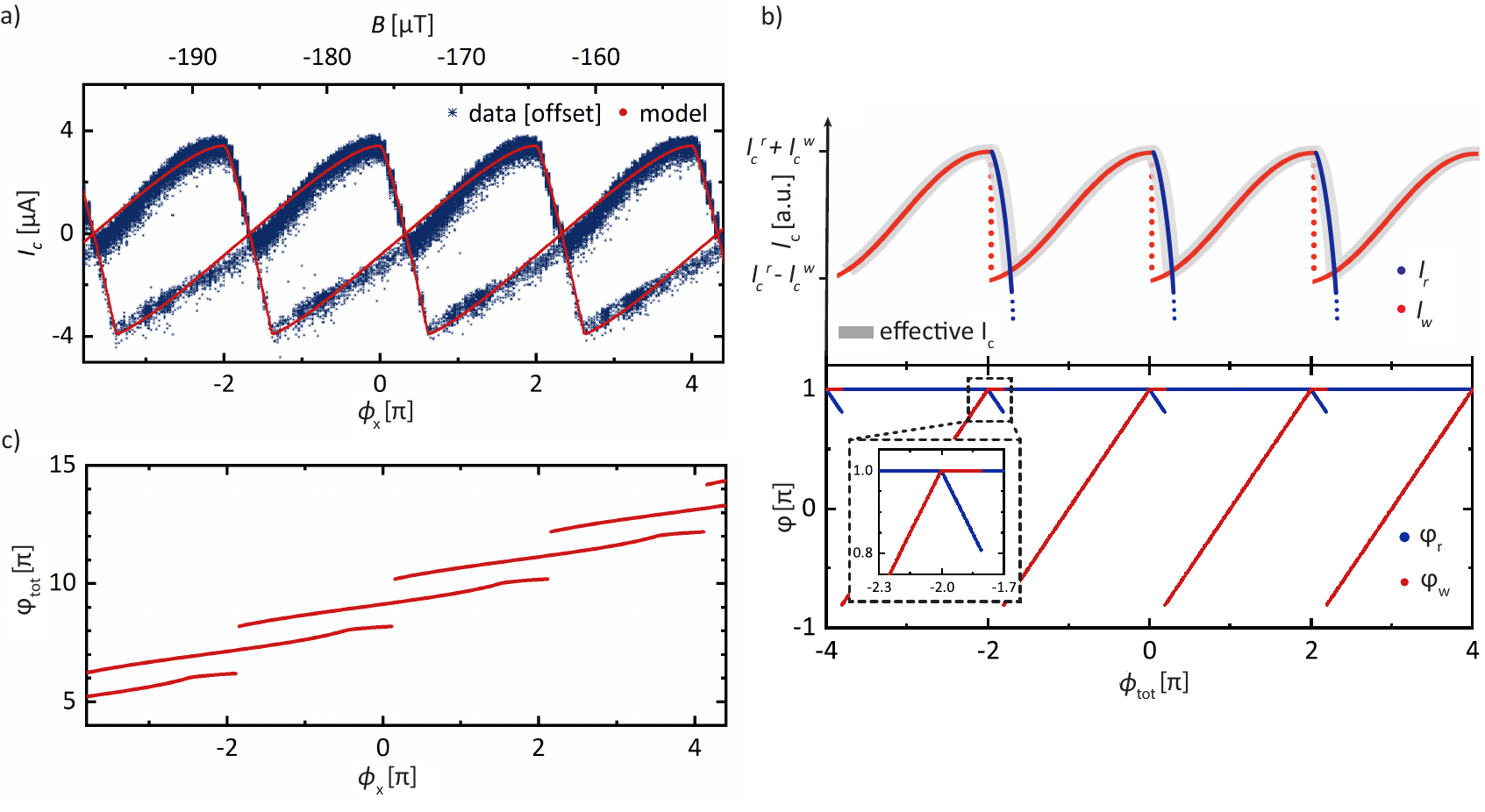}
    \caption{\textbf{SQUID oscillations and numerical model of the CPR.}
    a) Experimental data of the SQUID oscillations as function of external flux $\phi_{\text{x}}$ shown in blue.  The red data points are a fit to the data based on maximizing the critical current in the SQUID loop. Phase winding in the superconducting loop is responsible for the multivalued supercurrent. 
    b) Visual method to maximize $I_{\text{c}}$ as a function of total flux $\phi_{\text{tot}}$.  The upper panel plots the currents $I_{\text{r}}$ and $I_{\text{w}}$ for a short ballistic CPR with an amplitude ratio $I_{\text{c}}^{\text{r}}/I_{\text{c}}^{\text{w}}=$ 40.  The two currents evolve in opposite direction with increasing $\phi_{\text{tot}}$, due to $\varphi_{\text{r}}-\varphi_{\text{w}}=\phi_{\text{tot}}$. The resulting maximized $I_{\text{c}}$ is highlighted by a gray background and is composed of the weak junction and the reference junction CPR for the rising and falling side, respectively.  The lower panel shows the corresponding behavior of the junction phases $\varphi_{\text{r}}$ and $\varphi_{\text{w}}$, obtained through the numerical maximization model. The phases are mapped to the range of the CPR.  The switch between the observed weak and reference junction branch in the top panel is accompanied by a shift from a fixed $\varphi_{\text{r}}$ to $\varphi_{\text{w}}$ at flux values of multiple of $2\pi$.
    c) Numerically calculated $\phi_{\text{tot}}$ versus $\phi_{\text{x}}$ using the model described in the main text.  Inductance effects give rise to the multivalued $\phi_{\text{x}} (\phi_{\text{tot}})$, responsible for the intertwined branches visible in panel a).  
}%
    \label{fig:4 Data}
\end{figure*} 

Figure~\ref{fig:2}~b) presents the measured switching statistics of the SQUID in an extended magnetic field range at base temperature $T$ = \SI{30}{\milli \kelvin} of the cryostat. Visible is a multivalued $I_{\text{c}}$ that oscillates periodically around the offset current $I_{\text{c}}^{\text{r}}$ = \SI{160}{\micro \ampere} of the reference junction. Given the great difference in critical current amplitudes $I_{\text{c}}^{\text{r}}/I_{\text{c}}^{\text{w}} \sim 40$, we expect the SQUID to be highly asymmetric and the measured signal to reflect the CPR of the weak JJ. A second set of oscillations appears for field values from $B=$ \SI{-100}{\micro \tesla} upwards and is explained in Ref.~\cite{supp} by the behavior of the reference junction. The extracted oscillation period of a fixed parity branch, as defined in Fig.~\ref{fig:1}~a), is equal to $\delta B$ = \SI{23.2}{\micro \tesla} and therefore twice the value $\delta B= \Phi_{0}/A_{\circ}$ = \SI{11.1}{\micro \tesla} expected for the enclosed loop area $A_{\circ}$. 
The data represent striking resemblance to the two parity states of a topological JJ in the long junction limit with a $4\pi$ periodicity in flux. However, the amplitude of the signal deems this explanation unlikely. A single channel in the topological ballistic long junction can carry a current $I_{\text{c},4\pi}=E_{\text{Th}}e/\hbar = v_{\text{F}}e/l$ \cite{Beenakker2013}, with $E_{\text{Th}}$ being the Thouless energy, $v_{\text{F}}$ the Fermi velocity and $l$ the length of the weak link. $I_{\text{c}}$ in our junction would therefore have to be carried by at least $I_{\text{c}}^{\text{w}}l_{\text{w}}/(ev_{\text{F}}) \sim$ 116 perfectly ballistic channels  in parallel, assuming $v_{\text{F}}=$ \SI{3.09e5}{\meter \per \second} \cite{Li2017}. WTe$_2$ is expected to host a pair of hinge states on opposite edges of the crystal \cite{Choi2020}. Conducting states have been found to reside at step edges in the crystal where the number of vdW layers changes \cite{Kononov2020}, but we do not observe such crystal steps in optical microscopy for the used flake.  We conclude that it is unlikely for the current to be carried purely by ballistic hinge states. 

Instead, multivalued CPR measurements have previously been reported in devices containing a superconducting weak link with large kinetic inductance that is responsible for strong screening effects \cite{Murphy2017, Lefevre-Seguin1992, Friedrich2019, Hazra2014, Dausy2021}. In strong contrast to previous experiments, however, the device studied here contains a normal conducting weak link \cite{Endres2022, supp} and has negligible loop inductance, as will be shown later.  

The schematic in Fig.~\ref{fig:2}~c) introduces a set of additional inductances, $L_{\text{r}}$ and $L_{\text{w}}$,  that are placed in series to the reference and weak junction, respectively,  while potential mutual inductances are assumed to be negligible. In general,  the total phase
\begin{equation}
\phi_{\text{tot}} = \phi_{\text{x}}+ 2\pi(L_{\text{r}}I_{\text{r}}-L_{\text{w}}I_{\text{w}})/\Phi_{0},
\label{Equ:2}
\end{equation}
can differ strongly from the phase created by the external flux $\phi_{\text{x}}=2\pi \Phi_{\text{x}}/Phi_{0}$, due to the contribution induced by the currents $I_{\text{r}}$ and $I_{\text{w}}$ passing through the inductances in the SQUID arms, $L_{\text{r}}$ and $L_{\text{w}}$. We note, that while a screening current can distort the flux dependence of the critical current, it does not change its periodicity \cite{Tinkham1994}.

Deducing the CPR from the inductive SQUID measurement requires the knowledge of the phase dependences of $I_{\text{r}}(\phi_{\text{tot}})$ and $I_{\text{w}}(\phi_{\text{tot}})$ themselves.  In order to bypass this recursive constraint, we make an assumption about the CPRs of the JJs, that is based on the experimental data. We are then going to use this information in the next step to calculate $\varphi_{\text{r}}^{\text{max}}(\phi_{\text{tot}})$, for which $I_{\text{c}}(\phi_{\text{tot}})$ is maximized according to Eq.\ref{Equ:1}. Last, we are going to include screening effects in the model and obtain the relation $I_{\text{c}}(\phi_{\text{x}})$, which is placed in context to the experimental data. 

Our choice of CPR for the weak junction is based on two experimental observations. First, the rising slope of $I_{\text{c}}(\phi_{\text{x}})$ is non-linear, suggesting the same should hold for the CPR. Second, $I_{\text{c}}(\phi_{\text{x}})$ has self-crossings, implying that the phase of the reference junction does not remain fixed, contrary to the established expectation for a highly asymmetric SQUID. The behavior is possible, if the CPR of the weak junction contains abrupt changes, such as it is the case in the ballistic limit of the CPR. We illustrate this behavior later in the text and provide an additional discussion in \cite{supp}. $I_{\text{w}}(\varphi_{\text{w}})$ is modelled to be in the $2\pi$-periodic, ballistic short junction limit \cite{Bagwell1992}
\begin{equation}
I_w(\varphi_{\text{w}}) = I_{\text{c}}^{\text{w}} \frac{sin(\varphi_{\text{w}})}{\sqrt{1-sin^{2}(\varphi_{\text{w}}/2)}},
\label{equ: ballistic CPR}
\end{equation}
scaled by the amplitude of $I_{\text{c}}^{\text{w}}$.  The exact CPR of the reference junction plays little role in the further discussion, yet, since it is formed in the same material but with reduced length, we also model it as a short ballistic junction, scaled by $I_{\text{c}}^{\text{r}}$.  Independently, we have used length dependent measurements in Ref.~\cite{supp} and the analysis of the PdTe$_x$ diffusion profile in Ref.~\cite{Endres2022}, to verify that both junctions behave indeed as JJs and are not shorted by PdTe$_{\text{x}}$ \cite{Faucher2002niobium}. 

Having made this initial assumption, we continue to calculate the resulting $I_{\text{c}}$ by maximizing the current through the SQUID. The top panel in Fig.~\ref{fig:4 Data}~b) illustrates a visual method to maximize the critical current as a function of $\phi_{\text{tot}}$. The individual currents through the SQUID arms $I_{\text{r}}$ and $I_{\text{w}}$ follow Eq.~\ref{equ: ballistic CPR}. The currents evolve in opposite flux direction, due to the connection $\varphi_{\text{r}}-\varphi_{\text{w}}=\phi_{\text{tot}}$.
We start at the configuration $\varphi_{\text{r}}=\varphi_{\text{r}}^{\text{max}}$ and $\varphi_{\text{w}}=\varphi_{\text{w}}^{\text{max}}$, when the currents through the reference and weak junctions are at their maximum, $I_{\text{c}}^{\text{r}}$ and $I_{\text{c}}^{\text{w}}$, respectively.  
Moving from this point in negative direction of $\phi_{\text{tot}}$, $I_{\text{c}}$ follows $I_{\text{w}}$, plotted as a red curve. In the opposite direction towards positive values of $\phi_{\text{tot}}$, $I_{\text{w}}$ faces a sudden drop. Instead of following $I_{\text{w}}$, a higher $I_{\text{c}}$ is obtained by following $I_{\text{r}}$, drawn in dark blue, until the point when $I_{\text{r}}$ intersects with $I_{\text{w}}$. This creates a small flux range $\delta \phi_{\text{tot}} \propto \pi I_{\text{c}}^{\text{w}}/I_{\text{c}}^{\text{r}} \ll \pi$, in which $I_{\text{c}}$ is maximized by following $I_{\text{r}}$ rather than $I_{\text{w}}$, resulting in a changing $\varphi_{\text{r}}$, while $\varphi_{\text{w}}^{\text{max}}=\pi$ remains fixed. The situation is illustrated in the lower panel of Fig.~\ref{fig:4 Data}~b). Once the two current branches cross with evolving flux, $\varphi_{\text{r}}$ returns to its maximum value $\varphi_{\text{r}}^{\text{max}}=\pi$ and $I_{\text{c}}(\phi_{\text{tot}})$ follows the flux dependence of $I_{\text{w}}$. The well known behavior of the asymmetric SQUID is restored. While $\delta \phi_{\text{tot}}$ remains small in the above scenario, it can extend significantly in the experiment, due to the inductance effects, as described by Eq.~\ref{Equ:2}.

Given that $\varphi_{\text{r}}$ does not necessarily remain fixed in an asymmetric SQUID, we introduce next a numerical procedure to transfer the above model from the dependence of total flux $\phi_{\text{tot}}$ to the external flux $\phi_{\text{x}}$, the quantity applied in the experiment. The slope of $I_{\text{c}}(\phi_{\text{x}})$ is directly related to the inductance of the current carrying arm in the SQUID, via $dI_{\text{c}}/d\phi_{\text{x}} = \Phi_{0}/(2\pi (L_{\text{i}}+L_{\text{J}}^{\text{i}}))$, with $L_{\text{J}}^{\text{i}}$ being the Josephson inductance and $\text{i}=\text{r,w}$, depending on the considered data mapping the CPR of the reference or the weak junction. 

First,we numerically maximize the expression
\begin{equation}
I_{\text{c}}(\phi_{\text{tot}})= \max_{\varphi_{\text{r}}} \{ I_{\text{r}}(\varphi_{\text{r}}(\phi_{\text{tot}})) + I_{\text{w}}(\varphi_{\text{r}}(\phi_{\text{tot}})+\phi_{\text{tot}}) \}
\end{equation}
 with respect to $\varphi_{\text{r}}(\phi_{\text{tot}})$ for a given $\phi_{\text{tot}}$. $I_{\text{r}}$ and $I_{\text{w}}$ correspond to the currents through the respective SQUID arms, i.e. $I_{\text{r}} = I_{\text{c}}^{\text{r}}f(\varphi_{\text{r}})$ and $I_{\text{w}} = I_{\text{c}}^{\text{w}}f(\varphi_{\text{r}}+\phi_{\text{tot}})$. The two amplitudes $I_{\text{c}}^{\text{r}}=$ \SI{160}{\micro \ampere} and $I_{\text{c}}^{\text{w}}=$ \SI{3.75}{\micro \ampere} are determined from the fixed current background and the oscillation amplitude of the intertwined branches.
The lower panel in Fig.~\ref{fig:4 Data}~b) plots the obtained $\varphi_{\text{r}}$ and $\varphi_{\text{w}}$ in blue and red, respectively. Based on the choice of the CPR function in Eq.~\ref{equ: ballistic CPR}, $\varphi_{\text{r}}$ is mostly fixed at the maximum value $\varphi_{\text{r}}^{\text{max}}= \pi$, while $\varphi_{\text{w}}$ evolves linearly in $\phi_{\text{tot}}$, according to $\varphi_{\text{w}}=\phi_{\text{tot}}+\varphi_{\text{r}}^{\text{max}}$. However, a small range $\delta \phi_{\text{tot}}$ exists, where $\varphi_{\text{r}}$ changes in flux while $\varphi_{\text{w}}$ remains fixed, in agreement with the above introduced graphical method in the upper panel of Fig.~\ref{fig:4 Data}~b). 

Using $\varphi_{\text{r}}(\phi_{\text{tot}})$, it is now possible to extract the inductance effects and recalculate $\phi_{\text{x}}=\phi_{\text{tot}}-2\pi(L_{\text{r}}I_{\text{r}}-L_{\text{w}}I_{\text{w}})/\Phi_{0}$. Depending on the magnitude of the incorporated inductances and critical currents, self-inductance effects in the loop cause the connection $\phi_{\text{tot}}(\phi_{\text{x}})$ to become multi-valued, as it is visible from Fig.~\ref{fig:4 Data}~c). 

Finally, $I_{\text{c}}(\phi_{\text{x}})$ is plotted red in Fig.~\ref{fig:4 Data}~a).  Despite the long physical length $L_{\text{w}}$ = \SI{1.5}{\micro \meter} of the junction,  we find the bending of the gradually rising slope $dI_{\text{c}}/d\phi_{\text{x}}$ to be well reproduced by $f_{\text{w}}$ being in the $2\pi$ periodic, ballistic short junction limit. 
In both cases, the magnetic field dependencies of the current amplitudes are assumed to be negligible for the given field range.
Importantly, despite the great difference in critical current amplitudes of the embedded junctions, the model confirms that $\varphi_{\text{r}}$ does not remain fixed in flux. The experimental CPR of the SQUID is composed of the weak and the reference junction CPR. Even though the CPR of the weak junction is not necessarily uniquely in the short junction limit, it has to have a sharp transition in flux and therefore be close to ballistic in order to ease the shift between the fixed $\varphi_{\text{r}}$ and $\varphi_{\text{w}}$.  

Further, we extract the inductances $L_{\text{r}}$ = \SI{60}{\pico \henry} and $L_{\text{w}}$ = \SI{220}{\pico \henry}, by matching the rising and falling slope of the fit to the data.  An important result that distinguishes ours from previous experiments is that $L_{\text{r}}$ and $L_{\text{w}}$ by themselves exceed the sum of geometrical inductance $L_{\text{geo}} \approx$ \SI{27.0}{\pico H} \cite{Shatz2014} and kinetic inductance $L_{\text{kin}} \approx$ \SI{5.5}{\pico H} \cite{Annunziata2010} for the Nb SQUID loop. Possibly, additional JJs can form at the interface between the sputtered superconducting leads and the self-formed superconducting PdTe$_{\text{x}}$ \cite{Sinko2021},  yet given their critical current has to be larger than $I_{\text{c}}^{\text{r}}$, little inductance contribution is to be expected.  Instead, we attribute the origin of additional inductance and its asymmetry between the SQUID arms to the superconducting PdTe$_{\text{x}}$ that has self-formed at the interface between WTe$_2$ and Pd.  Further support of this interpretation is provided in Ref.~\cite{supp}, including the comparison of the data to different initial CPR assumptions. 

Finally, the multivalued $I_{\text{c}}$ can also be explained in the framework of excited vorticity states in an inductive SQUID \cite{Hopkins2005, Murphy2017,Hazra2014, Dausy2021, Friedrich2019}. Using the parameters obtained from our fit, we calculate the magnetic screening factor $\beta_{\text{L}} =\pi I_{\text{c}}^{\text{w}}(L_{\text{w}}+L_{\text{r}})/\Phi_{0} >$~1 \cite{Tinkham1994}, reflecting that an additional flux quantum can be created by the maximum circulating current through the weak JJ. The result is a multivalued $\phi_{\text{tot}}$ as a function of $\phi_{\text{x}}$,  as was shown in Fig.~\ref{fig:4 Data}~c) for given device parameters.  The above behavior can differ strongly even on a single sample chip. While the  second SQUID loop formed on the same WTe$_2$ flake (compare Fig.~\ref{fig:2}~a)) reveals the same behavior of the reference junction with multiple branches, we do not observe higher vorticity states in the SQUID oscillations. The absence of the feature is most likely connected to the overall smaller $I_{\text{c}}^{\text{w}}$, despite the shorter junction length with $l_{\text{w}}=$ \SI{1.2}{\micro \meter}.

\section*{Summary and Conclusion}

In conclusion, the established assumption of a fixed reference junction phase in flux does not hold for highly transparent junctions, even in the case of highly asymmetric critical current amplitudes. Furthermore, we have shown the complexity of subtle inductance effects that reach beyond the standard consideration of the loop inductance and might create misleading topological features.  It is therefore crucial for a correct CPR measurement to consider potential inductance contributions from the interfaces between the embedded junctions with the SQUID loop. The origin of such additional inductances can go beyond the here presented diffusion of PdTe$_x$ and may include defects implanted at the interface through various fabrication steps \cite{Annunziata2010, Niepce2019}. 
Despite these limitations, the here presented fitting routine allows to reproduce the experimental data closely. The best result was obtained by placing the weak junction in the short ballistic limit, as presented in Ref.~\cite{supp}. Our results establish WTe$_2$ as a promising platform for further experiment towards topological superconductivity.  

\textbf{Note.} Similar experimental results have been obtained recently in Ref.~\cite{Bernard2023}, however, the authors provide a different interpretation.

\section*{Acknowledgements}

This project has received funding from the European Research Council (ERC) under the European Union’s Horizon 2020 research and innovation programme: grant agreement No 787414 TopSupra, by the Swiss National Science Foundation through the National Centre of Competence in Research Quantum Science and Technology (QSIT), and by the Swiss Nanoscience Institute (SNI).
A.K. was supported by the Georg H.~Endress foundation.
D.M. and J.Y. acknowledge support from the U.S. Department of Energy (U.S.-DOE), Office of Science - Basic Energy Sciences (BES), Materials Sciences and Engineering Division.
H.S.A. was supported by the Gordon and Betty Moore Foundation's EPiQS Initiative through Grant GBMF9069 and the Shull Wollan Center Graduate Research Fellowship.
D.M. acknowledges support from the Gordon and Betty Moore Foundation’s EPiQS Initiative, Grant GBMF9069.
K.W. and T.T. acknowledge support from the Elemental Strategy Initiative conducted by MEXT, Japan and the CREST (JPMJCR15F3), JST and from the JSPS KAKENHI (Grant Numbers 19H05790 and 20H00354).

\section*{Author Contributions}
M.E. has fabricated the devices. M.E. and A.K. measured the devices. H.S.A., J.Y. and D.M. provided the WTe$_2$ crystals. K.W., T.T. provided the hBN crystals. M.E., A.K. and C.S.  analysed the data and wrote the manuscript. 

\section*{Competing Interests}

The authors declare no competing interest. 

\section*{Data availability}
All data in this publication are available in numerical form in the Zenodo repository \cite{Zenodo}.

\bibliography{bibliography_v2}

\end{document}


\title{Supplementary Information: Current-phase relation of a WTe$_2$ Josephson junction}%

\author{Martin Endres}
\affiliation{Department of Physics, University of Basel, Klingelbergstrasse 82, 4056 Basel, Switzerland}

\author{Artem Kononov}
\affiliation{Department of Physics, University of Basel, Klingelbergstrasse 82, 4056 Basel, Switzerland}

\author{Hasitha Suriya Arachchige}
\affiliation{Department of Physics and Astronomy, University of Tennessee, Knoxville, Tennessee 37996, USA}

\author{Jiaqiang Yan}
\affiliation{Department of Physics and Astronomy, University of Tennessee, Knoxville, Tennessee 37996, USA}
\affiliation{Material Science and Technology Division, Oak Ridge Laboratory,Oak Ridge, Tennessee 37831, USA}

\author{David Mandrus}
\affiliation{Department of Materials Science and Engineering, University of Tennessee, Knoxville, Tennessee 37996, USA}
\affiliation{Department of Physics and Astronomy, University of Tennessee, Knoxville, Tennessee 37996, USA}
\affiliation{Material Science and Technology Division, Oak Ridge Laboratory,Oak Ridge, Tennessee 37831, USA}

\author{Kenji Watanabe}
\affiliation{Research Center for Functional Materials, National Institute for Materials Science, 1-1 Namiki, Tsukuba 305-0044, Japan}

\author{Takashi Tanigutchi}
\affiliation{International Center for Materials Nanoarchitectonics, National Institute for Materials Science,  1-1 Namiki, Tsukuba 305-0044, Japan}

\author{Christian Sch{\"o}nenberger}
\affiliation{Department of Physics, University of Basel, Klingelbergstrasse 82, 4056 Basel, Switzerland}
\affiliation{Swiss Nanoscience Institute, University of Basel, Klingelbergstrasse 82, 4056 Basel, Switzerland}

\date{\today}%

\maketitle

\section{Materials and Methods}

\subsection*{Stacking of the vdW Heterostructure}

The following section describes the fabrication process of the devices, valid in general for both, singe JJs and SQUID loops. 
First,  the palladium (Pd) bottom contacts are fabricated on a p-doped silicon wafer with \SI{285}{\nano \meter} SiO$_2$ layer on top. This type of wafer is used throughout the complete fabrication process and is referred to as Si/SiO$_2$ wafer.  The structure is written via e-beam lithography, followed by metal deposition in an e-beam evaporator of \SI{3}{\nano m} titanium (Ti)  and \SI{15}{\nano m} Pd.  The lift-off of excess metal is done in warm acetone at \SI{50}{\degree C} for 1 hour. 
Next, the van-der-Waals materials are prepared. 
Hexagonal boron nitride (hBN) is exfoliated in ambient conditions on untreated Si/SiO$_2$ wafers using adhesive tape. Suitable flakes with a thickness between \SI{10}{\nano \meter} and \SI{30}{\nano \meter} are selected via optical contrast under the microscope \cite{Blake2007} and are checked to be uniform in dark-field mode. 
WTe$_{2}$ is an air sensitive van der Waals material prone to oxidation \cite{Ye2016, Hou2020, Liu2015}.  Starting from exfoliation until complete encapsulation, the crystals are handled inside a nitrogen filled glovebox with an oxygen level $<$ 1 ppm. We use flux grown, needle shaped WTe$_2$ crystals for exfoliation that are preferably oriented along their crystallographic a-axis \cite{Zhao2015}. 
The crystals are exfoliated using first an adhesive tape which is then pressed onto a polydimethylsiloxane (PDMS) stamp. The PDMS stamp is placed on a Si/SiO$_2$ substrate and the package is heated for 5 minutes at \SI{120}{\degree C} on a hotplate. After cool-down, PDMS is peeled off from the substrate and fitting flakes with thickness between \SI{10}{\nano \meter} and \SI{30}{\nano \meter} are selected using their optical contrast and dark field imaging.
We use a dry transfer technique \cite{Zomer2014} with a polycarbonate (PC)/PDMS stamp to successively pick up the selected hBN and WTe$_2$ flakes at $T$ = \SI{80}{\degree C} and release them on the patterned Pd bottom contacts by melting the PC at $T$ = \SI{150}{\degree C}.  Remaining PC residues are dissolved at room temperature for 1 hour in dichloromethane. 

\subsection*{Deposition of Superconducting Contacts}

Neither WTe$_2$ nor Pd are intrinsically superconducting and superconductivity only emerges when both materials are brought in contact to each other \cite{Endres2022,Kononov2021}. The phenomenon arises through the diffusion of Pd into the WTe$_2$ crystal and the formation of superconducting PdTe$_{\text{x}}$ \cite{Endres2022}. The external superconducting loops are fabricated using a recently developed recipe for high transparency junctions, as described in Ref.~\cite{Endres2022}.

First, an etch mask is written by e-beam lithography,  defining the later SQUID loop (or the contacts for single JJs) on top of the vdW stack.  In order to preserve the high contact transparency \cite{Endres2022}, the superconducting contacts are separated by a distance $l_{\text{Pd}} \approx$ \SI{0.5}{\micro \meter} from the JJ interface defined by the Pd bottom contact. We use a CHF$_{3}$/O$_{2}$ plasma in a reactive ion etcher (RIE) to etch through the encapsulating hBN layer into the WTe$_{2}$.  After the quick transfer to the sputtering machine, an argon (Ar) plasma at \SI{50}{W} is ignited for 1 minute.  Afterwards, \SI{100}{\nano \meter} Nb with a capping layer of \SI{3}{\nano \meter} platinum (Pt) capping layer are sputtered.  Alternatively, \SI{100}{\nano \meter} aluminium is deposited in an e-beam evaporator (also including a plasma etching step before).

\section{Counter measurement technique}

\begin{figure}[hp]
    \centering
    \includegraphics[width=0.65\linewidth]{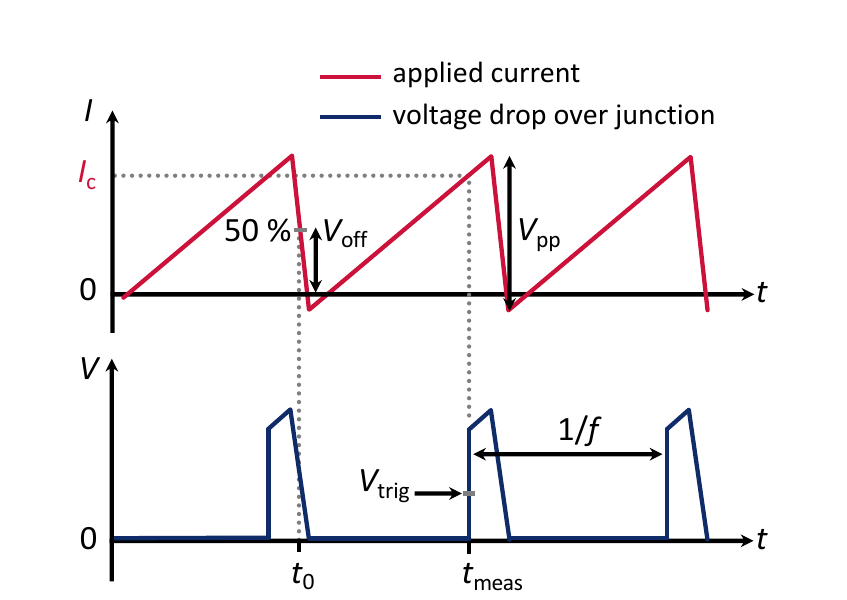}
    \caption{ \textbf{Counter measurement technique.} A sawtooth-shaped voltage $V_{\text{bias}}$ is applied to a resistor $R_{\text{bias}}$ in series with the device, effectively creating an alternating current $I$ (upper panel).  Simultaneously, the voltage $V_{\text{SQUID}}$ (lower panel) over the device is measured and a non-zero value is detected when $I>I_{\text{c}}$, the critical current of the SQUID. From the known input parameters of the ac-current it is possible to calculate the switching current.}%
    \label{fig:2 Counter}
\end{figure} 

The devices are probed in a quasi four-terminal configuration by sourcing an ac-current and monitoring the voltage drop over the SQUID.  We use the counter technique, outlined in Fig.~\ref{fig:2 Counter}, to measure $I_{\text{s}}$ and map the switching statistics of the device \cite{Indolese2020}.  
The sawtooth current $I$ is created by applying a sawtooth-shaped voltage $V_{\text{bias}}$ with frequency \textit{f} (\SI{5}{\hertz} -- \SI{1.7}{\kilo \hertz}), amplitude $V_{\text{pp}}$ and off-set voltage $V_{\text{off}}$ from zero (upper panel) to a resistor $R_{\text{bias}} \sim$ \SI{10}{\kilo \ohm} in series with the device. The voltage drop over the SQUID $V_{\text{SQUID}}$ (lower panel) is measured and forwarded to a counter.  For $I<I_{\text{c}}$, the device resides in a dissipationless state and $V_{\text{SQUID}}$ equals zero. Once $I=I_{\text{c}}$ is exceeded, transition to the resistive state sets in sharply and a voltage is detected. 
$I_{\text{c}}$ is calculated through 

\begin{equation}
I_{\text{c}} = \frac{1}{R_{\text{bias}}} \left( V_{\text{off}} + \frac{V_{\text{pp}}f}{\text{x}}(t_{\text{meas}}-\frac{1}{2f}\right),
\end{equation}

with the time $t_{\text{meas}}$,  taken between 50$\%$ of the falling slope of $V_{\text{bias}}$ and $V_{\text{SQUID}}$ exceeding a defined trigger level $V_{\text{trig}}$, when the junction switches into the resistive state.  $x$ denotes to the ratio of rising to falling slope per period $T=1/f$ of the drive signal.  The routine is repeated 200 times for every magnetic field value with all critical current values being recorded.

\section{Inductance effects in single Josephson junctions}

\begin{figure}[hp!]
    \centering
    \includegraphics[width=1\linewidth]{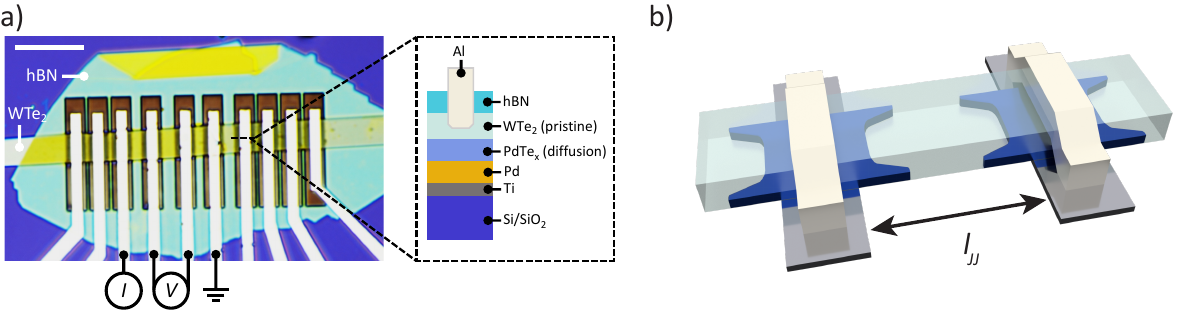}
    \caption{a) Optical image of multiple JJs formed in a WTe$_2$ crystal.  An illustration of the quasi four-terminal measurement scheme is illustrated at the bottom of the image. Visible is the elongated WTe$_2$ flake in horizontal direction, lying on vertically oriented Pd bottom contacts. The scale bar is \SI{10}{\micro \meter}. The fabricated layer sequence of the heterostructure across the black dashed line is shown on the right.  b) Illustration of a single Josephson junction with the effective diffusion profile of PdTe$_{x}$ highlighted in blue.}%
    \label{Supp:Length Dependence Sample}
\end{figure} 

\begin{figure}[hp!]
	\centering
	\includegraphics[width=1\linewidth]{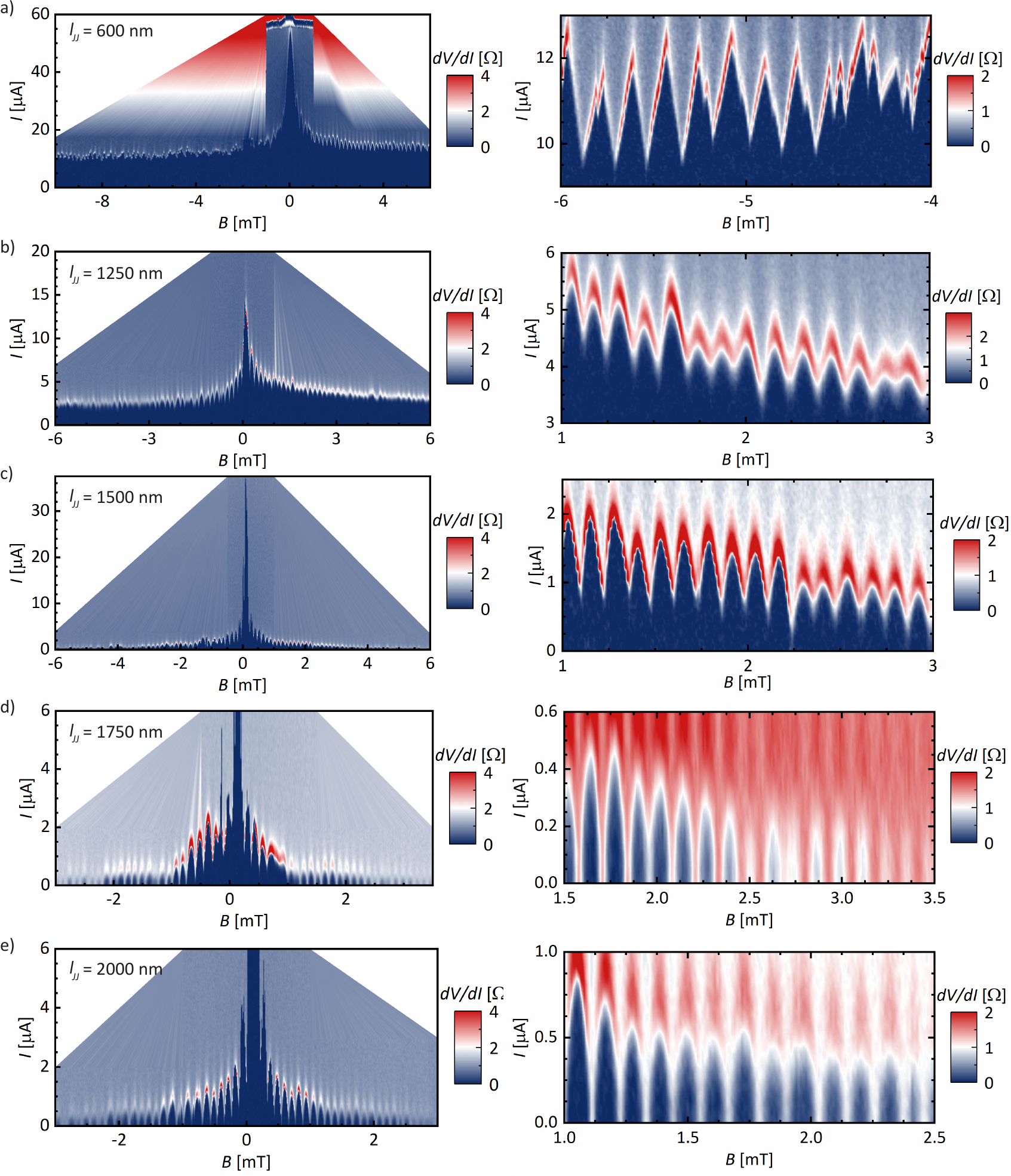}
	\caption{Differential resistance as a function of perpendicular magnetic field and applied bias current for different junction lengths $l_{\text{JJ}}$. The length of the JJs increases from panels a) through e), as labelled in the top left corner of the plots on the left. The plots in the right column are a zoom in to the data to their left. }%
	\label{Supp fig:Length Dependence Data}
\end{figure}

The complex switching behavior of the inductive SQUID can be simplified by investigating a single Josephson junction (JJ).  In the following section we present the study on JJs of different lengths. The governing transport regime can be shifted from being dominated by kinetic inductance to the Josephson inductance as the junction length increases. 

Fig.~\ref{Supp:Length Dependence Sample}~a) shows an optical image of the device with its multiple JJs on an elongated WTe$_2$ flake. The external superconducting contacts are formed by aluminium. A cross section through the device is shown on the right. The effective diffusion profile of PdTe$_{\text{x}}$ inside a single junction is illustrated in Fig.~\ref{Supp:Length Dependence Sample}~b).  

It has recently been shown that Pd diffuses inhomogeneously into the WTe$_2$ crystal, with greater extent along the edges compared to the bulk \cite{Endres2022}. The effective diffusion profile of the superconducting compound PdTe$_{\text{x}}$ is illustrated in blue in Fig.~\ref{Supp:Length Dependence Sample}~b). The inhomogeneous diffusion profile has extensive consequences for the electronic transport. First, the enhanced diffusion-length along the edges of WTe$_2$ could lead to an increased role of those edges in the Josephson transport. This is beneficial in case of topological edge/hinge states, since it would diminish the contribution from trivial bulk states. Second, the diffusion channels could form a nano-inductor with high kinetic inductance $L_{\text{K}} \propto l/(wd)$ \cite{Annunziata2010}, governed by the length $l$, width $w$ and thickness $d$ of the diffusion strip. 

We have fabricated a series of JJs with different spacing between the Pd bottom contacts, ranging from $l_{\text{JJ}}$ = \SI{600}{\nano \meter} to \SI{2}{\micro \meter}. DC-current measurements are performed in four-terminal configuration, as illustrated in Fig.~\ref{Supp:Length Dependence Sample}~a), with a standard lock-in technique at the base temperature of the cryostat $\sim$ \SI{30}{\milli K}.
Figure~\ref{Supp fig:Length Dependence Data} summarizes the differential resistance as a function of applied perpendicular magnetic field and current bias for the different junctions. Each row presents a junction length with the full range data shown on the left and a zoom-in of the data on the right. We begin with the shortest junction length $l_{\text{JJ}}$ = \SI{600}{\nano \meter} in Fig.~\ref{Supp fig:Length Dependence Data}~a). On a macroscopic scale, the critical current envelope, at which the junction switches from zero to finite resistance, peaks sharply around zero magnetic field and decays only slowly towards higher field values. A zoom in to the data, presented in the right panel of Fig.~\ref{Supp fig:Length Dependence Data}~a), reveals sharp sawtooth-like oscillations that are aperiodic in magnetic field.  The oscillation nodes are lifted from zero. The data resemble the characteristics of an asymmetric SQUID that is formed inside the WTe$_2$ junction by the spatially inhomogeneous PdTe$_{\text{x}}$ along the edges of the crystal. The asymmetry in critical current of the two junctions, $I_{\text{c}}^{\text{r}}/I_{\text{c}}^{\text{w}}$, can be quantified by the oscillation range in current amplitude between $I_{\text{c}}^{\text{low}}\sim$ \SI{9.7}{\micro \ampere} and $I_{\text{c}}^{\text{high}}\sim$ \SI{12.4}{\micro \ampere}.  We find $I_{\text{c}}^{\text{r}}/I_{\text{c}}^{\text{w}}= (I_{\text{c}}^{\text{high}}+I_{\text{c}}^{\text{low}})/(I_{\text{c}}^{\text{high}}-I_{\text{c}}^{\text{low}})\sim 8.1 \gg 1$ \cite{Kononov2020}, with $I_{\text{c}}^{\text{r}}$ and $I_{\text{c}}^{\text{w}}$ denoting the high and low critical current of the reference and weak junction, respectively.

With increasing junction length, shown in direction of the lower panels in Fig.~\ref{Supp fig:Length Dependence Data}, the overall shape of $I_{\text{c}}(B)$ resembles a ``Fraunhofer'' pattern of a single JJ in a magnetic field. In more detail, with increasing junction length, the oscillation amplitude reduces and the shape of the oscillations in magnetic field becomes more sinusoidal. Furthermore, the nodal offset from zero vanishes. 

In an asymmetric SQUID, such as it is the case for the shortest junction in Fig.~\ref{Supp fig:Length Dependence Data}~a), $I_{\text{c}}$ as a function of magnetic field maps the CPR of the weak JJ with smaller critical current \cite{DellaRocca2007}. The sawtooth-like CPR could be interpreted as being in the ballistic long junction limit, originating from the topological edge states of the material \cite{Murani2017}. We argue, however, that this interpretation is unlikely to be the case in the presented device when the aperiodicity and inconsistent change in amplitude of the oscillations are taken into account. Instead, the linear behavior in magnetic field is attributed to inductance effects in the junction \cite{Murphy2017, LefevreSeguin1992, Friedrich2019, Hazra2014, Dausy2021}. 

The evolution of $I_{\text{c}}(B)$ with junction length, from a sawtooth-like to a sinusoidal interference pattern, can be explained consistently in the framework of a SQUID with flux screening that is formed by the extended PdTe$_{\text{x}}$ arms along the WTe$_{2}$ crystal edges. We assume that the PdTe$_{\text{x}}$ diffusion along the edges is similar for all junctions lengths and gives a constant contribution to the kinetic loop inductance $L_{\text{K}}$. As the junction length increases, so does the ratio between Josephson inductance $L_{\text{J}}$ and kinetic inductance of the diffusion layer $L_{\text{K}}$:
\begin{equation}
\frac{L_{\text{J}}}{L_{\text{K}}} = \frac{\Phi_{0}}{2\pi L_{\text{K}} I_{\text{c}}}.
\label{ch7:equ:Lj/Lk}
\end{equation}
Longer junctions, due to their smaller value of $I_{\text{c}}$, have an increased Josephson inductance. If $L_{\text{J}}$ exceeds the kinetic inductance that originates from the PdTe$_{\text{x}}$ diffusion crystal, the phase difference over the junction will be determined by the external flux $\phi_{\text{x}}$ and a sinusoidal interference pattern is observed. 

\section{Verification of the inductance effects}

\begin{figure}[hp!]
	\centering
	 \includegraphics[width=11 cm]{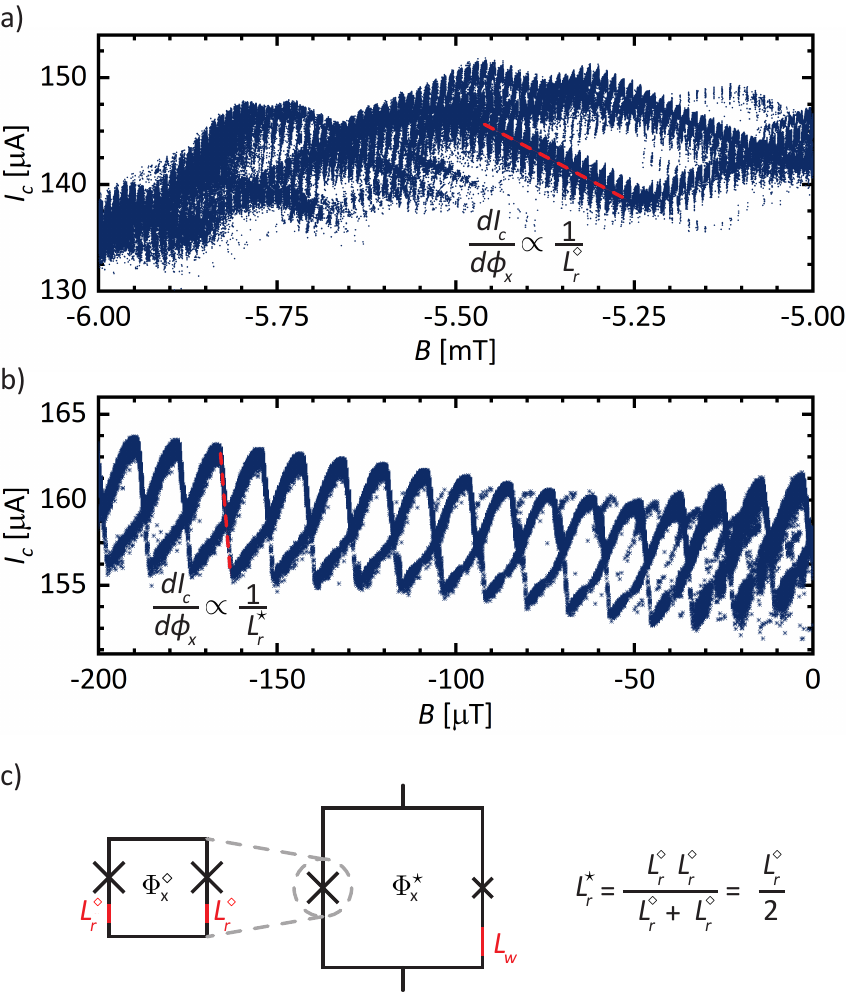}
	\caption{\textbf{Extraction of the reference junction series inductance $L_{\text{r}}$ for different magnetic field ranges through the linear slope of the oscillations. }a) Long range data of the SQUID that purely resembles the behavior of the reference junction. The SQUID oscillations are superimposed on top of the long range oscillations. b) High resolution measurement of the SQUID oscillations at a point in magnetic field. c) Schematic illustration to explain the factor two difference between the extracted slopes $L_{\text{r}}^{\diamond}$ and $L_{\text{r}}^{\star}$.}%
	\label{supp:fig: InductanceSlope}
\end{figure}

Consistently with the picture of inductance effects, the slope $dI_{\text{c}}/d\phi_{\text{x}}$, attributed to the phase change in the reference junction, yields the same inductance value for the long range data ($L_{\text{r}}^{\diamond}$) Fig.~\ref{supp:fig: InductanceSlope}~a) and the SQUID oscillations ($L_{\text{r}}^{\star}$) in Fig.~\ref{supp:fig: InductanceSlope}~b) and the long range data ($L_{\text{r}}^{\diamond}$) in Fig.~\ref{supp:fig: InductanceSlope}~b). We find $L_{\text{r}}^{\star} \sim $ \SI{81}{\pico \henry} and $L_{\text{r}}^{\diamond}\sim$ \SI{166}{\pico \henry} $\sim 2L_{\text{r}}^{\star}$. The factor of 2 arises, when the reference junction is considered a symmetric SQUID itself, due to the inhomogeneous PdTe$_{\text{x}}$ diffusion \cite{Endres2022}, as illustrated in Fig.~\ref{supp:fig: InductanceSlope}~c). In one case, the flux threads the area of the large Nb SQUID with an effective inductance $L_{\text{r}}^{\diamond}/2$, while in the other, it threads the reference junction area.  The good agreement of $L_{\text{r}}^{\diamond} \sim 2L_{\text{r}}^{\star}$ was obtained by using the center-to-center distance between the Nb contacts in the reference junction. The measure can be explained by the sample design, where the flux through half of the superconducting contact is screened into the junction \cite{Ghatak2018anomalous}.

\section{Comparison of CPR configurations in the fitting procedure}

\begin{figure}[hp!]
	\centering
	 \includegraphics[width=11 cm]{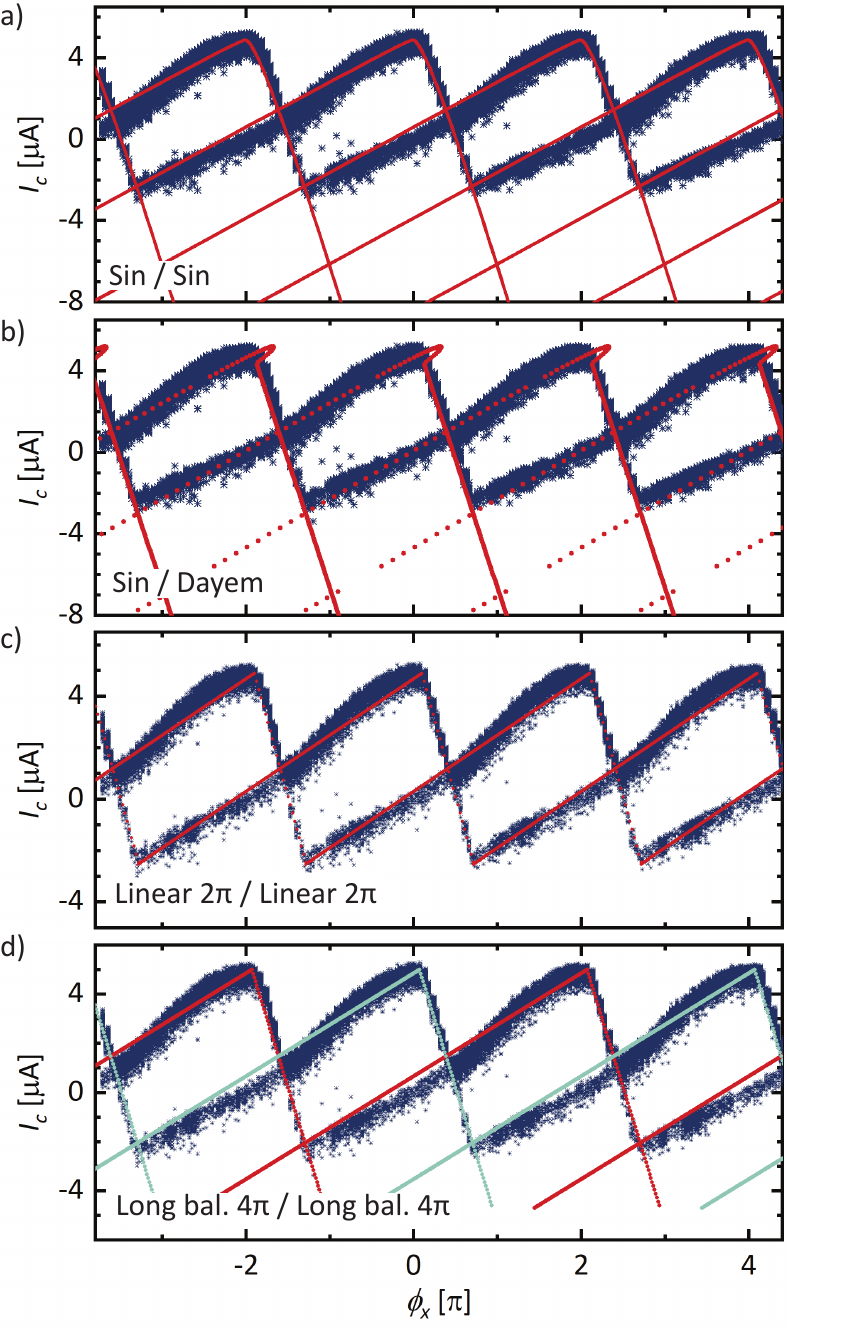}
	\caption{\textbf{Comparison of fit models with different CPR functions.} The weak junction and the reference junction, respectively, are assumed to have a) both a sinusoidal CPR,  b) a sinusoidal CPR and the behavior of Dayem bridge and c) both a linear CPR of an inductor. d) Topological junction in the ballistic long junction limit with a $4\pi$-periodic CPR.  The red and green branch correspond to two parity states that are shifted by a phase of $2\pi$ relative to each through quantum phase slips.}%
	\label{supp:fig: Fit model comparison}
\end{figure}

\begin{table}[]
\centering
\begin{tabular}{|c | c | c | c| c|}
\hline
$f_{w}$ & Sin($\varphi$) & Sin($\varphi$) & Linear $2\pi$ & Long bal. $4\pi$ \\
$f_{r}$ & Sin($\varphi$) & Dayem & Linear $2\pi$ & Long bal. $4\pi$ \\ \hline
$I_{c}^{w}$ [\SI{}{\micro \ampere}] & 60 & 10 & 3.8 & 5 \\ 
$L_{w}$ [\SI{}{\pico H}] & 450 & 400 & 200 & 80 \\ 
$I_{c}^{r}$ [\SI{}{\micro \ampere}] & 104 & 154 & 160 & 160 \\ 
$L_{r}$ [\SI{}{\pico H}] & 270 & - & 80 & 80 \\ \hline
\end{tabular}
\caption{\textbf{Summary of device parameters used in the fits of Fig.~\ref{supp:fig: Fit model comparison}.}  $f_{\text{w}}$ and $f_{\text{r}}$ denote to the normalized CPR function of the weak and reference junction used in the fit model. In case of the reference junction being modelled as Dayem bridge, all inductances are incorporated through the slope $1/L = dI/d\phi_{\text{x}}$ of the CPR. ``Long bal. $4\pi$'' is the abbreviation for a $4\pi$ periodic topological JJ in the ballistic long junction limit.}
\label{supp:tab:fit parameters}
\end{table}

In the main text of the manuscript, we have fitted the data under the assumption that both junctions in the SQUID are in the short ballistic limit. The assumption was based on the non-linearity of $I_{\text{c}}(\phi_{\text{x}})$ in the rising slope and the fact that the phase $\varphi_{\text{r}}$ of the reference junction does not remain fixed and therefore requires an abrupt jump in the CPR function. 

In Fig.~\ref{supp:fig: Fit model comparison} we present additional fits to the data $I_{\text{c}}{\phi_{\text{x}}}$ of the main text, based on different CPR functions of the two junctions.  Details of fit parameters are summarized in Tab.~\ref{supp:tab:fit parameters}.

Figure~\ref{supp:fig: Fit model comparison}~a) addresses both, the weak junction and the reference junction a sinusoidal CPR function. In order to obtain the instability of the reference junction phase, $I_{\text{c}}^{\text{w}}$ has to be heavily increased compared to the observed data range.  The amplitude of the reference junction is adjusted such that $I_{\text{c}}^{\text{r}}=$\SI{164}{\micro \ampere}$-I_{\text{c}}^{\text{w}}$. The increased $I_{\text{c}}^{\text{w}}$  gives rise to additional excited SQUID states that extend below the data. The series inductances of both SQUID arms, $L_{\text{w}}$ and $L_{\text{r}}$ are adjusted such that the fit matches the slope of the data for a chosen $I_{\text{c}}^{\text{w}}$. In this case it is not possible to determine one single parameter set that fits the data. Reason for this is, that the limited data range of the measurement is not enough to picture higher excited branches with very low switching probability. 

Following a similar argument, Fig.~\ref{supp:fig: Fit model comparison}~b) assumes a sinusoidal CPR for the weak junction, while treating the reference junction as Dayem bridge. The approach would hold, if PdTe$_{\text{x}}$ had created a superconducting short in the reference junction. Here,  all inductance effects are included in the slope of the CPR, according to $1/L_{\text{r}}=dI_{\text{c}}/d\phi_{\text{x}}$, resulting in a periodicity that exceeds $2\pi$. Visible around the maximum $I_{\text{c}}$ is an overshoot of the fit, that originates from the maximization process of $I_{\text{c}}$. Due to the shallow slope of the Dayem bridge, $\varphi_{\text{w}}$ continues to evolve in flux beyond the maximum value of $I_{\text{w}}=I_{\text{c}}^{\text{w}} Sin(\phi_{\text{r}}+\phi_{\text{x}})$ until the point when $I_{\text{w}}<I_{\text{r}}$.

In a third option, plotted in Fig.~\ref{supp:fig: Fit model comparison}~c), both CPR functions are considered to be of a linear inductor due to the dominating inductance effects. 

Last, Fig.~\ref{supp:fig: Fit model comparison}~d) considers the weak and reference junction to be in the topological regime. Both junctions are modelled to be in the ballistic long junction limit with its characteristic sawtooth CPR. It should be noted, that the effective fit does not change significantly, if the reference junction is considered to be in the ballistic short junction limit, as can be seen from the presented fits above. The red and blue branch of the fit correspond to the Majorana bound states with opposite parity, shifted by a phase of $2\pi$ relative to each.  The extension of the fit below the fit originates form the included inductance $L_{\text{w}}=L_{\text{r}}=$ \SI{80}{\pico \henry} in the SQUID arms.  

In conclusion,  none of the four alternative scenarios matches the experimental data in all its features. In particular the curvature of the rising slope is best reproduced by the both junctions being in the short ballistic limit, as presented in the main text. 

\section{Instability of the reference junction phase $\varphi_{r}$}

We argue in the following section that current crossings in a multivalued $I_{\text{c}}(\phi_{\text{x}})$ directly imply that the reference junction phase $\varphi_{\text{r}}$ in an asymmetric SQUID can not remain fixed at its maximum value. 

Let us assume that the initial situation is as observed in Fig.~3 a) of the main text, such that $I_{\text{c}}(\phi_{\text{x}})$ has self-crossings. Each crossing would be connected to two values in total flux, called $\phi_{\text{tot},1}$ and $\phi_{\text{tot},2}$, respectively. If $\varphi_{\text{r}}$ remained fixed and only $\varphi_{\text{w}}$ evolved, the device would be analogue to an ac-SQUID, containing only a single JJ.  The current $I$ circulating in such a device is described by $I=I_{\text{c}} f(\phi_{\text{tot}})$, with $I_{\text{c}}$ being the critical current of the JJ and $f$ being its CPR. The total flux in the ac-SQUID with inductance $L$ is given by $\phi_{\text{tot}}=\phi_{\text{x}}+I(\phi_{\text{tot}})L$. Since $I$ and $\phi_{\text{x}}$ are identical at the crossing point, the two equations directly imply that $\phi_{\text{tot},1} = \phi_{\text{tot},2}$. For this reason it is not possible to obtain self-crossings in $I_{\text{c}}(\phi_{\text{x}})$ if only one junction phase evolves.

\section{Frequency Dependence of the Switching Statistics}

\begin{figure}[hp!]
   \centering
   \includegraphics[width=11 cm]{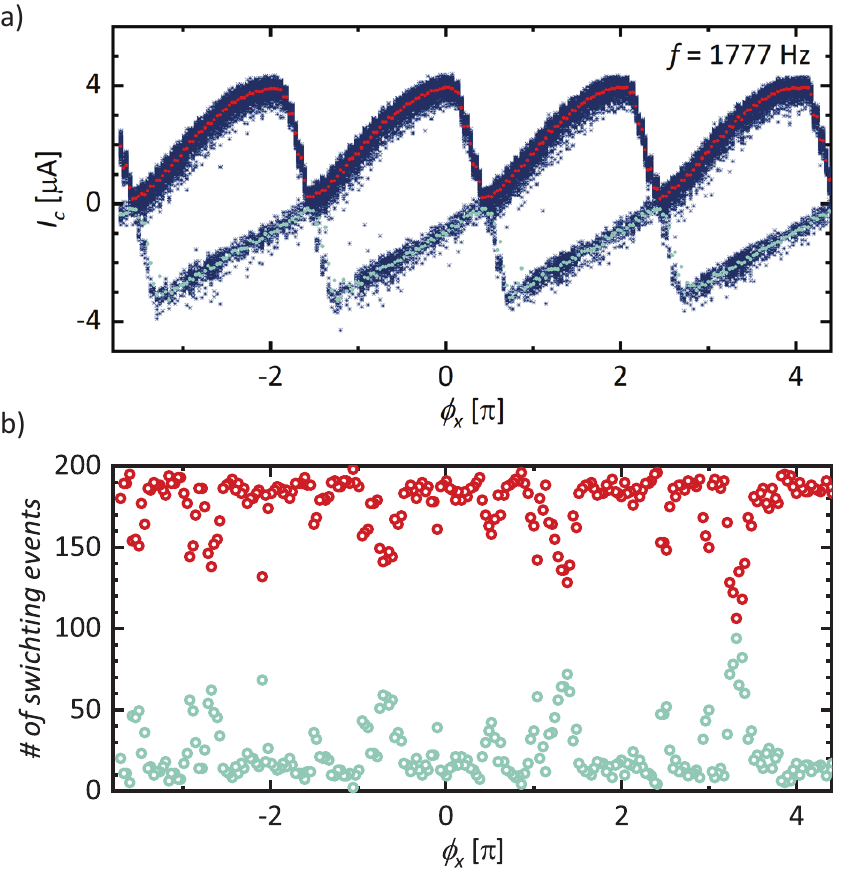}
    \caption{\textbf{Occupation of the zeroth and higher vorticity states at applied ac frequency $f=$ \SI{1777}{\hertz}.} a) SQUID oscillations with color-coded average value for the zeroth and higher vorticity state above and below the current crossing point. b) Number of switching events for the two intertwined branches. Each magnetic field step contains 200 switching events, of which $\sim 13\%$ are occupying the higher vorticity state below the crossing point of $I_{\text{c}}$ values.}%
    \label{supp:fig: Branch occupation}
\end{figure} 

\begin{figure}[hp!]
   \centering
   \includegraphics[width=11 cm]{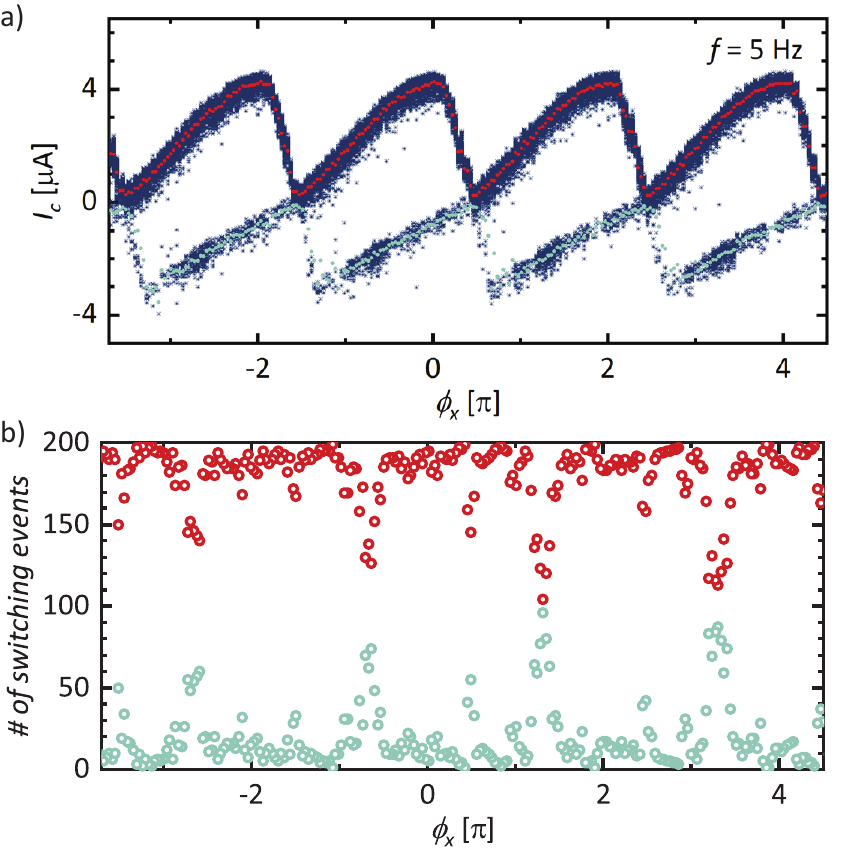}
    \caption{\textbf{Occupation of the zeroth and higher vorticity SQUID states at applied ac frequency $f=$ \SI{5}{\hertz}.} a) SQUID oscillations with color-coded average value for the zeroth and higher-order state above and below the current crossing point. b) Number of switching events for the two intertwined branches. Each magnetic field step contains 200 switching events, of which $\sim 11\%$ are occupying the higher vorticity state below the crossing point of $I_{\text{c}}$ values.}%
    \label{supp:fig: Frequency Dependence}
\end{figure} 

The formation of higher vorticity states in the SQUID can be understood in terms of a  a particle with mass moving in a two-dimensional SQUID potential, as described by Ref.~\cite{LefevreSeguin1992}. This two-dimensional potential for an inductive DC SQUID can be seen as an analogue to the standard washboard potential of a single JJ \cite{Tinkham1994}. 
In the following discussion we introduce the nomenclature of a ground state and higher vorticity state of the SQUID to describe the data \cite{LefevreSeguin1992}. These terms denote the $I_{\text{c}}$ values above and below the crossing point of the multivalued $I_{\text{c}}(\phi_{\text{x}})$ measurement and relate to the circulating current in the system. 
Figure~\ref{supp:fig: Branch occupation}~a) plots the average values of the switching distributions for the ground state and the higher vorticity state in red and turquoise, respectively. The frequency of the applied ac-current is $f=$ \SI{1777}{\hertz}. Following the same color code, Fig.~\ref{supp:fig: Branch occupation}~b) traces the number of switching events in each of the states.  At each magnetic field value, the critical current is measured 200 times in total and the switching events are distributed amongst the two visible states. The mean values of the ground state and higher vorticity state are $\sim$ 177.3 and $\sim$ 22.7 switching events per magnetic field step, relating to a ratio $r \sim$ 0.128. 
We observe the higher vorticity SQUID state down to the lowest applied sawtooth-frequency $f=$ \SI{5}{\hertz} in Fig.~\ref{supp:fig: Frequency Dependence}~a). In this case, the mean value of the higher-order state is reduced to $\sim$ 19.4 switching events per magnetic field value, resulting in $r \sim$ 0.107.

\bibliography{bibliography_v2_supplm}